\documentclass[pra,aps,floatfix,amsmath,superscriptaddress,
twocolumn]{revtex4}

\usepackage{amssymb}
\usepackage{graphicx}
\usepackage{graphics}
\usepackage{amsmath}
\usepackage{amsthm}
\usepackage{color}
\usepackage{dsfont}

\usepackage{mathtools}
\def\ra{\rangle}
\def\la{\langle}

\newtheorem{theorem}{Theorem}

\newcommand{\bea}{\begin{eqnarray}}
\newcommand{\eea}{\end{eqnarray}}
\newcommand{\be}{\begin{equation}}
\newcommand{\ee}{\end{equation}}
\newcommand{\ba}{\begin{equation}\begin{aligned}}
\newcommand{\ea}{\end{aligned}\end{equation}}

\newtheorem{definition}{Definition}
\theoremstyle{remark}

\def\be{\begin{equation}}
\def\ee{\end{equation}}

\newcommand{\mH}{\mathcal{H}}

\newcommand{\mS}{\mathcal{S}}

\newcommand{\lr}{\rangle\langle}

\newcommand{\tr}{{\rm Tr}}

\newcommand{\mbb}[1]{\mathbb{#1}}

\def\>{\rangle}
\def\<{\langle}

\begin{document}
	
	\preprint{APS/123-QED}
	\title{Any entanglement of assistance is polygamous}

	\author{Yu Guo}
	\email{guoyu3@aliyun.com}
	\affiliation{Institute of Quantum Information Science, Shanxi Datong University, Datong, Shanxi 037009, China}
	\affiliation{School of Mathematics and Statistics, Shanxi Datong University, Datong, shanxi 037009, China}

	\begin{abstract} 
		We propose a condition for a measure of quantum correlation to be polygamous without the traditional polygamy inequality. It is shown to be equivalent to
		the standard polygamy inequalities for any continuous measure of quantum correlation with the polygamy power.
		We then show that any  
		entanglement of assistance is polygamous but not monogamous and any faithful entanglement measure is not polygamous.
	\end{abstract}
	
	\maketitle

	
\section{Introduction}

Monogamy law of quantum correlations as one of the most striking features
in quantum world has been explored extensively ever since the distribution
of three qubit state entanglement discovered by Coffman, Kundu,
and Wootters (CKW)~\cite{Coffman,Gour2005pra,Osborne,Gour,Ouyongcheng2007pra2,Zhuxuena2014pra,Bai,Choi,Luo,Kumar,Kim2009,Buscemi,Kim2009pra,Lizongguo,Streltsov,Kim2012pra,Braga,Liusiyuan,Reid,Oliveira2014pra,Regula2014prl,Eltschka,Zhuxuena2015pra,Lancien,Lami,Song,Luo2016pra,Kumar2016pra,Jia,Kim2016pra,Chengshuming,Allen}.
A monogamy relation is quantitatively displayed as an inequality of the following form:
\be\label{basic}
Q(A|BC)\geq Q(A|B)+Q(A|C)\;,
\ee
where $Q$ is a measure of bipartite quantum correlation and $A,B,C$ are three subsystems of a composite quantum system,
the vertical bar indicates the bipartite split across which we will measure the (bipartite) entanglement.
Dually,
the polygamy relation in literature is expressed as 
\be
Q(A|BC)\leq Q(A|B)+Q(A|C)\;.
\label{polygamydefinition}
\ee
However, Eq.~(\ref{basic}) [resp. Eq.~(\ref{polygamydefinition})] captures only partially 
the property that $Q$ is monogamous (resp. polygamous). For example, it is well known that if $Q$ does not satisfy 
these relations, it is still possible to find a positive power $\alpha\in\mbb{R}_{+}$, 
such that $Q^\alpha$ satisfies the relation. 	
It was shown that many other measures of entanglement 
satisfy the monogamy relation~(\ref{basic}) or polygamy relation~(\ref{polygamydefinition}) if $Q$ is replaced by $Q^\alpha$ for some $\alpha>1$~\cite{Coffman,Gour2005pra,Osborne,Gour,Ouyongcheng2007pra2,Zhuxuena2014pra,Bai,Choi,Luo,Kumar,Kim2009}.

By now, we only know that entanglement of assistance and entanglement of assistance associated with the Tsallis-$q$ entropy are polygamous for any multipartite systems \cite{Buscemi,Kim2012pra,Kim2016pra,Kim2010pra}.
Concurrence of assistance, tangle of assistance and negativity of assistance are proved to be polygamous only for multiqubits systems \cite{Gour2005pra,Gour,Kim2009,Lizongguo}.
The polygamy problem for all other entanglement measures remain open for high dimensional systems.

Very recently, we 
introduced a definition of monogamy without inequalities in~\cite{GG}, which capture the nature of monogamy relation completely.
In this paper we present a condition of polygamy without inequalities as in Eq.~(\ref{polygamydefinition}). 
Our definition of polygamy (see Definition~1 below) does not present as a polygamy relation.
But, as what we will show, our definition is consistent with the traditional notion of polygamy, by showing that $Q$ is polygamous according to our definition if and only if there exists an $\beta>0$ such that $Q^\beta$ satisfy~(\ref{polygamydefinition}). 
Consequently, any entanglement of assistance is monogamous according to our definition.
We also prove that any faithful entanglement measure (a measure is said to be faithful if it is zero only on separable states) is
not polygamy and any entanglement of assistance is not monogamous.

Throughout this paper, a measure of quantum correlation $Q$ refers to any quantity that characterizes the ``quantumness'' contained in bipartite quantum systems, such as
entanglement, quantum discord~\cite{Ollivier,Henderson}, 
measurement-induced nonlocality~\cite{Luoshunlong}, 
quantum deficit~\cite{Oppenheim}, and other quantum correlations introduce recent years~\cite{Wiseman,WPM2009,Groisman,Luo2008,Guowu2014srep,Guo2015ijtp}, etc.
We denote by $\mS(\mH^{ABC})\equiv\mS^{ABC}$ the set of density matrices acting on a tripartite Hilbert space $\mH^{A}\otimes\mH^B\otimes\mH^C\equiv\mH^{ABC}$ associated with some tripartite system. 
For $\rho^{ABC}\in\mS^{ABC}$, $\rho^{AB}\equiv\tr_C\rho^{ABC}$ and $\rho^{AC}\equiv\tr_B\rho^{ABC}$ denote the marginal states on $\mH^{AB}$
and $\mH^{AC}$, respectively.

\section{An equivalent definition of polygamy}

\begin{definition}\label{main}
	Let $Q$ be 
	a measure of quantum correlation.
	We call
	$Q$ is \textit{polygamous} if for any $\rho^{ABC}\in\mathcal{S}^{ABC}$ that satisfies 
	\be\label{defcond}
	Q(\rho^{A|BC})>\max\{Q(\rho^{AB}),Q(\rho^{AC})\}>0\;,
	\ee
	we have 
	$\min\{Q(\rho^{AB}),Q(\rho^{AC})\}>0$.
\end{definition}

Definition~\ref{main} does not involve the standard polygamy relation as~(\ref{polygamydefinition}). However, we can derive a more quantitative inequality
from~(\ref{defcond}):

\begin{theorem} Let $Q$ be a continuous measure of quantum correlation.
	Then, $Q$ is pologamous according to Definition~\ref{main} if and only if 
	there exists $0<\beta<\infty$ such that
	\be\label{power}
	Q^\beta(\rho^{A|BC})\leq Q^\beta(\rho^{AB})+Q^\beta(\rho^{AC})
	\ee
	holds for any $\rho^{ABC}\in\mS^{ABC}$ with fixed $\dim\mH^{ABC}=d<\infty$.
\end{theorem}

\begin{proof}
	For any given $\rho^{ABC}$, we assume that $Q(\rho^{A|BC})=x$, $Q(\rho^{AB})=y$
	and $Q(\rho^{AC})=z$. 
	If $x\leq y$ or $x\leq z$, Eq.~(\ref{power}) is obvious.
	If $x>\max\{y,z\}>0$, then $y>0$ and $z>0$ since $Q$ is polygamous.
	Therefore there always exists $\gamma>0$ such that 
	\be \label{f}
	1\leq \left(\frac{y}{x}\right)^\gamma+\left( \frac{z}{x}\right)^\gamma\;,
	\ee 
	since $\left( \frac{y}{x}\right)^\gamma \rightarrow 1$ 
	and $\left( \frac{z}{x}\right)^\gamma\rightarrow 1$ when $\gamma$ decreases.
	Let $f(\rho^{ABC})$ be the largest value of $\gamma$ that saturates the inequality~(\ref{f}).
	Since $f$ is continuous and $\mS^{ABC}$ is compact,
	we get
	\be\label{beta}
	\beta\equiv\inf\limits_{\rho^{ABC}\in\mS^{ABC}}f(\rho^{ABC})<\infty\;,
	\ee 
	which satisfies Eq.~(\ref{power}).
\end{proof}

Note that
the original polygamy relation of $Q^\beta$ can be preserved when we lower the power~\cite{Kumar}:
Let $\rho^{ABC}\in\mS^{ABC}$ and $Q$ be a polygamy measure of quantum correlation.
Then $Q^\beta(\rho^{A|BC})\leq Q^\beta(\rho^{AB})+Q^\beta(\rho^{AC})$
implies $Q^\gamma(\rho^{A|BC})\leq Q^\gamma(\rho^{AB})+Q^\gamma(\rho^{AC})$
for any $\gamma\in[0,\beta]$.
(Note that, in~\cite{Kumar}, the bipartite correlation measure $Q$
is assumed to be normalized. This condition however, is not necessary,
which can be easily checked following the argument therein.)
We thus call $\beta$ defined in Eq.~(\ref{beta}) the \emph{polygamy power} of $Q$, i.e.,
$\beta(Q)$ is the supremum for $Q^{\beta(Q)}$ satisfies
Eq.~(\ref{power}) for all the states.  
For the case of $Q$ is entanglement assistance $E_a$ (see Eq.~(\ref{EoA}) below) associated with some entanglement 
measure $E$, it is always 
continuous (see Proposition in Appendix). Therefore $\beta(E_a)$ exists for $E_a$.
We list some examples in Table~\ref{tab:table}.
Polygamy power $\beta(E_a)$ is a dual concept of the monogamy power $\alpha(E)$ (see Theorem 1 in~\cite{GG}). 
$\beta(E_a)$ may depend also on the dimension $d\equiv\dim\mH^{ABC}$ as that of $\alpha(E)$~\cite{GG}.
Now, the pair $[\alpha(E),\beta(E_a)]$
reflects the share-ability of continuous entanglement measure $E$ completely. 
This pair of power indexes advances our understanding of 
multipartite entanglement
although these quantities are difficult to calculate.
\begin{table}
	\caption{\label{tab:table}The comparison of the polygamy power of several 
		assisted entanglement: concurrence of assistance $C_a$, negativity of assistance $N_a$, entanglement of assistance ${E_f}_a$ associated with the entanglement of formation $E_f$, tangle of assistance $\tau_a$ and entanglement of assistance ${T_q}_a$ associated with the Tsallis-$q$ entropy measure $T_q$.}
	\begin{ruledtabular}
		\begin{tabular}{cccc}
			$E_a$& $\beta(E_a)$ & System & Reference \\ \colrule
			$C_a$ & $2$& $2^{\otimes 3}$ & \cite{Gour,Gour2005pra}\footnotemark[1]$^,$\footnotemark[2]\\
			$N_a$& $\geq 2$ & $2^{\otimes n}$ &\cite{Kim2009}\footnotemark[1]\\
			${E_f}_a$& $\geq 1$ & any systems & \cite{Buscemi,Kim2012pra}\\
			$\tau_a$& $\geq 1$ & $2^{\otimes n}$ &\cite{Lizongguo}\\
			${T_q}_a$, $q\ge1$ &   $\geq 1$    & any system & \cite{Kim2010pra,Kim2016pra}\\
		\end{tabular}
	\end{ruledtabular}
	\footnotetext[1]{For pure states.}
	\footnotetext[2]{ $\beta(C_a)\geq2$ is proved in~\cite{Gour,Gour2005pra}. The equality $\beta(C_a)=2$ follows from the saturation by $|\psi\ra^{AB}\otimes|\psi\ra^C$ as in Eq.~\eqref{eg3} below.}
\end{table}

\section{Results}

We show below that any faithful entanglement measure is not polygamous
while any entanglement of assistance is polygamous. 
Recall that, for any given bipartite entanglement measure $E$ (for pure states),
the corresponding entanglement of assistance is defined by~\cite{DiVincenzo,GG}
\be\label{EoA}
E_a(\rho^{AB})=\max_{\{p_i,|\psi_i\ra\}}\sum_ip_iE(|\psi_i\ra)\;,
\ \rho^{AB}\in\mathcal{S}^{AB}\;.
\ee

\begin{theorem}\label{puretomix} 
	Let $E_a$ be an entanglement of assistance as above.
	If $E_a$ is polygamous (according to Definition~\ref{main}) on pure tripartite states 
	in $\mH^{ABC}$, then it is also polygamous on
	mixed sates acting on $\mH^{ABC}$.
\end{theorem}

\begin{proof}
	Let $\rho^{ABC}=\sum_{j}p_j|\psi_j\lr\psi_j|^{ABC}$ be a tripartite mixed state acting on $\mH^{ABC}$ with $\{p_j,|\psi_j\ra^{ABC}\}$ being the optimal decomposition such that
	\be
	E_a(\rho^{A|BC})=\sum_{j}p_jE_a\left(|\psi_j\ra^{A|BC}\right)\;.
	\ee
	We also assume w.l.o.g. that $p_j>0$.
	We now suppose $E_a(\rho^{A|BC})>E_a(\rho^{AB})>0$, $E_a(\rho^{AB})>E_a(\rho^{AC})$, and denote $\rho_{j}^{AB}\equiv\tr_{C}|\psi_j\lr\psi_j|^{ABC}$.
	Note that 
	\be
	E_a\left(|\psi\ra^{A|BC}\right)=E\left(|\psi\ra^{A|BC}\right)\geq E_a\left(\rho^{AB}\right)
	\ee 
	for any $|\psi\ra^{ABC}\in \mH^{ABC}$ since any ensemble of $\rho^{AB}$
	can be obtained from $|\psi\ra^{ABC}$ by some LOCC and LOCC can not increase entanglement on average [note that for mixed state $\rho^{A|BC}$, it is unknown whether $E_a(\rho^{A|BC})\geq E_a(\rho^{AB})$ in general].
	On the other hand,
	\be\nonumber
	\sum_{j}p_jE_a\left(|\psi_j\ra^{A|BC}\right)>E_a(\rho^{AB})\geq \sum_{j}p_jE_a\left(\rho_{j}^{AB}\right).
	\ee
	It follows that there exists some $j_0$ such that $E_a\left(|\psi_{j_0}\ra^{A|BC}\right)>E_a\left(\rho_{j_0}^{AB}\right)$ for some $j_0$. If $E_a\left(\rho_{j_0}^{AB}\right)>0$, then
	$E_a\left(\rho_{j_0}^{AC}\right)>0$ since $E_a$ is polygamous on pure tripartite states by assumption.
	We thus get
	\be
	E_a(\rho^{AC})\geq\sum_{j}p_jE_a\left(\rho_{j}^{AC}\right)>0\;.
	\ee
	If $E_a\left(\rho_{j_0}^{AB}\right)=0$, then $\rho_{j_0}^{AB}=\rho_{j_0}^A\otimes\rho_{j_0}^B$ with $\rho_{j_0}^B$
	is a pure state (note that $\rho_{j_0}^A$ is not pure, or else, $E(|\psi_{j_0}\ra^{A|BC})=0$, a contradiction).
	Therefore $|\psi_{j_0}\ra^{ABC}\cong |\psi\ra^{AC}|\psi\ra^B$, which reveals that
	$E_a(\rho_{j_0}^{AC})=E(|\psi\ra^{AC})=E(|\psi_{j_0}\ra^{A|BC})>0$.
	This completes the proof.  	
\end{proof}

\begin{theorem} 
	For any entanglement measure $E$, $E_a$ is polygamous.
\end{theorem}

\begin{proof} 
	We only need to show 
	$E_a$ is polygamous for tripartite pure states by Theorem~\ref{puretomix}.
	Note that $E_a(|\psi\ra^{A|BC})\geq E_a(\rho^{AB})$
	and $E_a(|\psi\ra^{A|BC})\geq E_a(\rho^{AC})$
	for any pure state $|\psi\ra^{A|BC}$, where
	$\rho^{AB}$ and $\rho^{AC}$ are the reduced states.
	If $E_a(|\psi\ra^{A|BC})> E_a(\rho^{AB})>0$,
	then $\rho^{AC}$ a mixed state and the ranks of $\rho^A$ and $\rho^C$ are at least 2.
	Without loss of generality, we suppose 
	$\rho^{AC}=\sum_{j=1}^2\lambda_j|x_j\ra|y_j\ra\la x_j|\la y_j|$,
	where $|x_1\ra$ and $|x_2\ra$ are linearly independent,
	$|y_1\ra$ and $|y_2\ra$ are linearly independent.
	Then $\rho^{AC}$ can be rewritten as
	$\rho^{AC}=|\psi_1\ra\la\psi_1|+|\psi_2\ra\la\psi_2|$,
	where $|\psi_1\ra=cd|x_1\ra|y_1\ra+e|x_2\ra|y_2\ra$ and
	$|\psi_2\ra=d|x_1\ra|y_1\ra-ce|x_2\ra|y_2\ra$ 
	(here, $|\psi_{1,2}\ra$ is unnormalized) with
	$(1+c^2)d^2=\lambda_1$ and $(1+c^2)e^2=\lambda_2$.
	This reveals $E_a(\rho^{AC})>0$.
	By Definition~\ref{main}, $E_a$ is polygamous on pure states, which completes the proof together with Theorem~\ref{puretomix}.
\end{proof}

\begin{theorem}\label{Eisnotpoly}
	Any faithful entanglement measure can not be polygamous.
\end{theorem}

Theorem~\ref{Eisnotpoly} is obvious from the following two examples.
The first example is the the generalized GHZ-class state that admit the multipartite Schmidt decomposition~\cite{Guoyu2015jpa,Guoyu2015qip}.	
\\

\noindent{\bf Example 1}
~We consider the following pure state
\be\label{ghz-class}
|\psi\ra^{A|B_1B_2\cdots B_n}
=\sum_{j=0}^{k-1}\lambda_j|j^{(0)}\rangle|j^{(1)}\rangle\otimes\cdots\otimes|j^{(n)}\rangle\;,
\ee
where $\{|j^{(0)}\rangle\}$ is an orthonormal
set in $\mH^A$, and $\{|j^{(i)}\rangle\}$ is an orthonormal set of $\mH^{B_i}$,
$\sum_j\lambda_j^2=1$, $\lambda_j >0$,
$k\leq\min\{\dim\mH^A,\dim\mH^{B_1},\dots,\dim\mH^{B_n}\}$,
$i=1$, 2, $\dots$, $n$, $n\geq 3$.
We always have $E(|\psi\ra^{A|B_1B_2\cdots B_n})>0$ while $E(\rho^{AB_i})=0$ 
for all $1\leq i\leq n$.
\\

\noindent{\bf Example 2}
~For the three qubit state
\be\label{eg2}
|\phi\ra^{ABC}
=\frac{1}{\sqrt{5}}(\sqrt{2}|000\ra+\sqrt{2}|110\ra+|111\ra)\;,
\ee
we can easily calculate that
$C(|\phi\ra^{A|BC})\approx0.9798$, $C(\rho^{AB})\approx0.7999$
and $\rho^{AC}$ is separable.

For $2\otimes 2\otimes 2^{m}$ systems, $m\geq 1$, it is shown in~\cite{Luo}
that 
$N^\beta(|\psi\ra^{A|BC})\leq N^\beta(\rho^{AB})+N^\beta(\rho^{AC})$
and 
$E^\beta(\rho^{A|BC})\leq E^\beta(\rho^{AB})+E^\beta(\rho^{AC})$
for $E$ is the convex-roof extended negativity or $E=E_f$ whenever $\beta\leq0$.
However, these statements are not true from the states in Eqs.~(\ref{ghz-class},\ref{eg2}).

By Definition 1 in~\cite{GG}, an entanglement measure $E$ is monogamous if for any $\rho^{ABC}\in\mathcal{S}^{ABC}$ that satisfies the disentangling condition
\be\label{cond}
E(\rho^{A|BC})=E(\rho^{AB})>0	
\ee
we have that $E(\rho^{AC})=0$.  
We remark here that, if $E_a(\rho^{A|BC})=E_a(\rho^{AB})>0$ and $E_a(\rho^{AC})$=0,
then 
\be\label{eg3}
\rho^{ABC}=|\psi\ra^{AB}\la\psi|\otimes\rho^C
\ee 
for some entangled pure state $|\psi\ra^{AB}\in\mH^{AB}$ and $\rho^C$ is of rank one.
If $\rho^C$ in Eq.~(\ref{eg3}) is a mixed state, then $E_a(\rho^{A|BC})=E_a(\rho^{AB})>0$ and $E_a(\rho^{AC})>0$.
That is, there exits mixed state $\rho^{ABC}$ that does not satisfy the disentangling condition~(\ref{cond}).
We thus obtain the following:

\begin{theorem}
	Any assistance of entanglement is not monogamous.
\end{theorem}

\section{Conclusions and discussions}

In conclusion, we improved the traditional definition of 
polygamy for measures of quantum correlation. 
For continuous measure $Q$, our definition is equivalent to the polygamy relations in the form of~\ref{polygamydefinition} with $Q$ is replaced by $Q^\beta$. 
We completely settled the polygamy 
problem for the 
case of entanglement by showing that 
any entanglement measure is not polygamous and
any entanglement of assistance is always polygamous but not monogamous.
Our results covers all the previous results about the polygamy of entanglement.
Along this line, it remains to determine the polygamy power of entanglement assistance,
which need to be further explored.
In addition, we should note that for any given measure of quantum correlation,
it cannot be monogamous and polygamous simultaneously 
in general. The only states that both monogamous and polygamous are
those in~(\ref{eg3}) whenever $\rho^C$ is pure.

\begin{acknowledgements}
	This work is supported by
	the Natural Science Foundation of Shanxi Province
	under Grant No. 201701D121001 and the National Natural
	Science Foundation of China under Grant No. 11301312.
\end{acknowledgements}

\bibliographystyle{spmpsci}      

\appendix

\section*{Appendix}

In~\cite{Guo2013qip} we proved that $C$ is a continuous function.
We show below that, for any entanglement measure $E$, the associated entanglement of assistance $E_a$ is continuous provided that $E$ is continuous on pure states. With no 
loss of generality, we consider
the case where $E$ is concurrence. The other cases, i.e., 
$E=N$, $\tau$, $E_f$, $T_q$, or the R\'{e}nyi-$\alpha$ entanglement $R_{\alpha}$, etc., can be argued similarly.\\

\noindent\textbf{Proposition.} {\it $C_a$ is continuous for
	both finite- and infinite-dimensional systems, i.e.,
	\be
	\lim_{n\rightarrow\infty}C_a(\rho_n) =C_a(\rho)\ {\rm
		whenever}\ \lim_{n\rightarrow\infty}\rho_n=\rho\ \
	\ee
	in the trace-norm topology.}

\begin{proof} To prove the continuity of $C_a$, let us extend
	the concurrence of  states to that of self-adjoint trace-class
	operators.

	Let $A$ be a self-adjoint trace-class operators acting on
	$\mH^A\otimes \mH^B$. We define the concurrence of assistance of $A$ by
	\begin{eqnarray}
	C_a(A)={\rm Tr}(|A|)C_a(\frac{|A|}{{\rm Tr}(|A|)})\;,
	\end{eqnarray}
	where $|A|=(A^\dag A)^{\frac{1}{2}}$.
	It is immediate that $C_a(A)\leq C_a(B)$ whenever $0\leq |A|\leq
	|B|$.
	In fact, for any rank-one projection decomposition of $|A|$,
	$|A|=\sum_i\lambda_i|\psi_i\rangle\langle\psi_i|$, $\lambda_i>0$,
	then for any rank-one projection decomposition of $|B|$,
	$|B|=\sum_i\lambda_i|\psi_i\rangle\langle\psi_i|
	+\sum_j\delta_j|\phi_j\rangle\langle\phi_j|$, $\delta_j>0$,
	where $\sum_j\delta_j|\phi_j\rangle\langle\phi_j|=|B|-|A|$. 
	This leads to $C_a(A)\leq C_a(B)$.
	
	Assume that $\rho_n,\rho\in {\mathcal S}^{AB}$ and  
	$\lim_{n\rightarrow\infty}\rho_n=\rho$.
	Let $\vartheta_n=\rho-\rho_n$ and let
	\begin{eqnarray*}\vartheta_n
		=\sum_{k(n)}\lambda_{k(n)}|\eta_{k(n)}\rangle\langle\eta_{k(n)}|
	\end{eqnarray*}
	be its spectral decomposition.
	
	We claim that
	\be
	C_a(\rho)=C_a(\rho_n+\vartheta_n)\leq C_a(\rho_n)-C_a(\vartheta_n)\;.
	\ee
	For any  $\varepsilon>0$, there exist ensembles $\{p_{k(n)}$,
	$|\psi_{k(n)}\rangle\}$ and $\{q_{l(n)}$, $|\phi_{l(n)}\rangle\}$ of
	$\rho_n$ and ${|\vartheta_n|}$, respectively,  and
	$0<\epsilon_1,\epsilon_2< \varepsilon$, such that
	\begin{eqnarray*}
		C_a(\rho_n)=\sum_{k(n)}p_{k(n)}C(|\psi_{k(n)}\rangle)+\frac{\epsilon_1}{2}
	\end{eqnarray*}
	and
	\begin{eqnarray}
	C_a({|\vartheta_n|})
	=\sum_{l(n)}q_{l(n)}C(|\phi_{l(n)}\rangle)+\frac{\epsilon_2}{2}.
	\label{1}
	\end{eqnarray}
	It follows that
	\begin{eqnarray*}
		&&C_a(\rho_n+\vartheta_n)\geq C_a(\rho_n-|\vartheta_n|)\\
		&\geq&\sum_{k(n)}p_{k(n)}C(|\psi_{k(n)}\rangle)
		-\sum_{l(n)}q_{l(n)}C(|\phi_{l(n)}\rangle)\\
		&=&C_a(\rho_n)-C_a(\vartheta_n)-\frac{\epsilon_1-\epsilon_2}{2}\;.
	\end{eqnarray*}
	Since $\varepsilon$ is arbitrarily given, the claim is proved.

	Similarly, using $C_a(\rho_n)=C_a(\rho-\vartheta_n)\geq
	C_a(\rho-|\vartheta_n|)$, we  obtain
	\begin{eqnarray*}
		C_a(\rho_n)\geq C_a(\rho)-C_a(\vartheta_n)\;,
	\end{eqnarray*}
	which, together with Eq.~(\ref{1}), implies that
	\begin{eqnarray*}
		|C_a(\rho_n)-C_a(\rho)|\leq C_a(\vartheta_n)\;.
	\end{eqnarray*}
	Observing that $C_a(\vartheta_n)\rightarrow0$ $(n\rightarrow\infty)$
	since $C_a(\vartheta_n)\leq\sum_{k(n)}\sqrt{2}|\lambda_{k(n)}|$ and
	${\rm Tr}(|\vartheta_n|) =\sum_{k(n)}|\lambda_{k(n)}|\rightarrow0$,
	we get $\lim_{n\rightarrow\infty}C_a(\rho_n)=C_a(\rho)$ as desired.
\end{proof}

The discussion above implies that, if $E(|\psi\ra^{AB})=f(\rho^A)$ is 
a continuous function of $\rho^A$,
where $|\psi\ra^{AB}\in \mH^A\otimes \mH^B$, $\rho^A={\rm Tr}_B|\psi\ra^{AB}\la\psi|$,
then $E_a$ is continuous as well.


\end{document}